\documentclass[aps, pra,twocolumn,groupedaddress,showpacs,nofootinbib]{revtex4-1}
\usepackage{amsmath, amsthm, amssymb}
\usepackage{amscd}
\usepackage{setspace}
\usepackage{graphicx}
\usepackage{color} 
\usepackage{dcolumn}

\begin{document}

\title{Phenomenological noise model for superconducting qubits: \\
two-state fluctuators and 1/f noise}
\author{Dong Zhou}
\author{Robert Joynt}
\affiliation{Department of Physics, University of Wisconsin-Madison, Wisconsin
53706, USA}

\begin{abstract}
We present a general phenomenological model for superconducting qubits
subject to noise produced by two-state fluctuators whose couplings to 
the qubit are all roughly the same.  In flux qubit
experiments where the working point can be varied,  it is possible to
extract both the form of the noise spectrum and the number of fluctuators. 
We find that the noise has a broad spectrum consistent with 1/f noise and
that the number of fluctuators with slow switching rates is surprisingly small:
less than $100$.  
If the fluctuators are interpreted as unpaired surface spins, 
then the size of their magnetic moments is surprisingly large. 
\end{abstract}
\date{\today}


\maketitle

\section{introduction}

Superconducting qubits based on Josephson junctions are promising candidates for 
quantum information processing \cite{Clarke08,Makhlin01}.  
Integrated-circuit fabrication technologies provides
a relatively straightforward route to scale up the number of qubits, {
and the qubit coherence times have been prolonged dramatically since the 
superconducting charge, phase and flux qubit designs were first developed 
over a decade ago
\cite{Nakamura1997,*Bouchiat1998,*Nakamura1999,*Vion2002,
Yu2002,*Martinis2002,*Berkley2003,
Mooij1999,*vanderWal2000,*Friedman2000,*Chiorescu2003,
Paik2011}.
However, detailed mechanism of decoherence due to the 
coupling of the Josephson device to external noise sources is 
still not fully understood \cite{Clarke08}.
}

Recent experiments on superconducting qubits show that 1/f flux noise is an
important source of decoherence \cite{Yoshihara06,Kakuyanagi07,Wellstood87}.
Experiments over the years have agreed on certain universal characteristics of this 
noise: 
(1) it has weak dependence on a wide range of parameters such as 
SQUID loop geometry, inductance, material, etc.;
(2) it has an approximately 1/f noise power spectrum and the magnitude ranges from 
$0.01-100(\mu\Phi_0)^2/Hz$ at the frequency $1 Hz$, where $\Phi_0=h/2e$ is the 
magnetic flux quantum \cite{Wellstood87,Foglietti86,Bialczak07, Yoshihara06,Kakuyanagi07}.  

The origin of this low-frequency noise at milli-Kelvin temperature has been a puzzle 
for over $20$ years and is still under active debate \cite{Koch07,Faoro08,deSousa07}.
There are indications that a high density of unpaired surface spins on the SQUIDs 
may be the physical causes of the noise \cite{Bialczak07,Koch07,deSousa07,
Sendelbach08,Yoshihara2010,Gustavsson2011,Sendelbach09}.
These defect sites behave as two-state fluctuators that switches between their 
two states due to thermal activations and/or other interactions.

In this paper, we present a phenomenological model of the fluctuators. 
The physical parameters of the model can be extracted from qubit measurements at 
different working points.
Analysis of experiments \cite{Yoshihara06,Kakuyanagi07}
produces estimations of the effective magnetic moment and noise power spectrum 
density that are comparable to the experimental findings.
Our chief new result is that the number of slow fluctuator is small, 
less than $100$ and possible even of order $10$.  

The paper is organized as follows. In Sec. \ref{sec:model}, we describe and solve 
the model. This gives results for free induction decay (FID), energy relaxation (ER)
and spin echo (SE) signals.  In Sec. \ref{sec:fit}, we summarize
our assumptions for the flux qubit systems and demonstrate how to extract the 
physical parameters of the model from experimental data. In Sec. \ref{sec:conc} we 
discuss the results.

\section{noise model}
\label{sec:model}

The superconducting flux qubits consist of a superconducting loop with three 
Josephson junctions \cite{Mooij1999,*vanderWal2000,*Chiorescu2003}. 
The two relevant states are the clockwise and counter-clockwise 
persistent current states in the loop and the loop is effectively a quantum 
two-level system.
The Hamiltonian of the superconducting flux qubit can be written as \cite{Clarke08}
\begin{align}
H_{qb} = -\frac{\varepsilon}{2}\sigma_z-\frac{\Delta}{2}\sigma_x-\frac{1}{2}h(t)\sigma_z,
\end{align}
where $\varepsilon$ and $\Delta$ are the energy difference and tunneling splitting
(Josephson coupling) between the clockwise and counter-clockwise current states, 
$h(t)$ is the flux noise in the environment and $\sigma_{x,y,z}$ are 
the Pauli matrices.
The energy difference is proportional to the applied flux through the superconducting
loop
\begin{align}
\varepsilon = 2I_p\left(\Phi_\text{ext}- {\Phi_0/2}\right),
\end{align}
where $I_p$ is the persistent current and {
    $\Phi_\text{ext}$ is the externally applied magnetic flux in the loop.
    When $\Phi_\text{ext}$ is half a flux quantum, the two current states are 
    degenerate in energy.
}
The flux noise $h(t)$ is described by a time-dependent classical field.
The eigenenergy of the qubit is thus
\begin{align}
B_0 = \sqrt{\varepsilon^2+\Delta^2}.
\end{align}
The angle $\theta=\tan^{-1}(\Delta/\varepsilon)$ is related to the working point of the
device: $\theta=\pi/2$ is the optimal point and  $\theta=0$ is the pure dephasing 
point.

The flux noise is induced by an ensemble of fluctuators, all fluctuating independently, 
giving rise to random telegraph noise (RTN).
Assuming a total number of $K$ fluctuators, the Hamiltonian can be written  in the 
following form after a basis transformation
\begin{align}
H = -\frac{1}{2} B_0\sigma_z -\frac{1}{2}\sum_{k=1}^K s_k(t)\vec{g}_k\cdot \vec{\sigma}.
\end{align}
Here  we redefine the $z$-axis to be the eigenenergy axis.
${g_k}$ is the coupling constant of the $k$'th fluctuator.
Note all fluctuators have the same $\theta$ value since flux fluctuation is along the 
$\varepsilon$ direction.
$s_k(t)$ is the random time sequence due to the $k$'th fluctuator and switches between
 the two values $-1$ and $1$ with an average switching rate $\gamma_k$.
For a single fluctuator, the noise auto-correlation function is 
\begin{align}
\overline{s(t)s(t')}=\exp(-2\gamma|t-t'|).
\end{align}
the power spectrum is given by 
\begin{align}
S_{\text{RTN}}(\omega)=&\frac{g^2}{2\pi}\int_{-\infty}^{\infty}
\overline{s(0)s(t)}e^{i\omega t}dt \notag \\
=&\frac{1}{2\pi}\frac{4\gamma g^2}{\omega^2+4\gamma^2}.
\end{align}
As is well-known, an ensemble of fluctuators with $1/\gamma$ distribution of their 
switching rates gives rise to $1/f$ noise power spectrum \cite{Kogan08}.

With the criteria 
\begin{align}
g_k \cos\theta \begin{matrix}
<\\
>\end{matrix}
\gamma_k 
\label{eq:cri}
\end{align}
we can put the fluctuators into two categories, the fast ones ($<$) and slow ones ($>$).
$K = M+N$ where $M$ ($N$) is the number of slow (fast) fluctuators.
The fast and slow fluctuators have qualitatively different effects on the qubit
time evolution \cite{Joynt_2009,Zhou10PRA, Zhou10QIP,Paladino02,
Galperin06}. Fast is synonymous with weakly-coupled or Markovian, as can be seen
from Eq. \ref{eq:cri}. The fast fluctuators can be treated with Redfield theory
and they give rise to exponential decay of phase 
coherence. {
On the other hand, slow is synonymous to strongly-coupled or non-Markovian
and Redfield theory cannot be applied.
In general, for classical Markovian noise or Gaussian noise, the dephasing rates
can be related to the noise spectral density and filter functions 
\cite{Slichter96,Martinis03,Cywinski08}.
A list of filter functions for common pulsing sequences can be found in Table
I of Ref. \cite{Cywinski08}.
}

    In this classical noise model, decoherence is a result of averaging the unitary
    time evolutions over all the possible noise sequencies $s_k(t)$.
The quasi-Hamiltonian method allows us to carry out this averaging analytically 
and treat the fast and slow fluctuators on an equal footing \cite{ Joynt_2009,Zhou10PRA}.
The qubit dynamics is described by a transfer matrix acting on the qubit
Bloch vector, i.e., $\vec{n}(t) = T(t) \vec{n}(t=0)$,  
while the transfer matrix is generated by a non-Hermitian quasi-Hamiltonian.
In the case of a single qubit interacting with a single fluctuator, 
the quasi-Hamiltonian has the form 
\begin{equation*}
H_{q}=-i\gamma +i\gamma \tau _{1}+\left[ B_{0}L_{z}+\tau _{3}\otimes \vec{g}
\cdot \vec{L}\right] ,
\end{equation*}
    where $\tau_{i}$ are Pauli matrices associated with the fluctuator, 
    and $L_i$ are
    the $SO(3)$ generators associated with the qubit Bloch vector.
    Note the classical two-valued fluctuating field is mapped into a spin-$1/2$ 
    particle in this formalism.
    The transfer matrix is given by 
$T(t)=\left\langle x_{f}\right\vert \exp (-iH_{q}t)\left\vert i_{f}\right\rangle$
where $\left\vert i_{f}\right\rangle =\left\vert x_{f}\right\rangle =[1;1]/\sqrt{2}$
correspond to unbiased fluctuator.
Exact diagonalization of $H_q$ is possible only for $\theta=0$ while perturbation 
expansion can be used in general to calculate the transfer matrix $T(t)$.

Signals from common experimental pulsing protocols, such as energy relaxation (ER),
Hahn spin echo (SE) and free-induction (FID), can be calculated with the 
quasi-Hamiltonian method as well \cite{Zhou10PRA}.
    For these pulsing schemes, the qubit is initially in the ground state and the 
    probability of the qubit being in the excited state is measured at time $t$.
    In the ER scheme, a single $\pi$ pulse is applied at the beginning of the 
    measurement.
    In the FID scheme, two $\pi/2$ pulses are applied, one at the beginning  and 
    the other at the end.
    The SE scheme has the two $\pi/2$ pulses as in the FID scheme and another $\pi$
    pulse in the middle of the time evolution, i.e., $t/2$.
For our qubit-fluctuators model, the pulsed signals are given by 
\begin{align}
n_\text{ER}(t) \simeq&  e^{-(2\sum_m\gamma_m\epsilon_{2m}^2\sin^2\theta+\Gamma_{1})t}\\
n_\text{SE}(t) \simeq& e^{-(\Gamma_2+\Gamma_3)t}\left[1 + \sum_{m=1}^{M}\epsilon_{1m}\sin(g_m\cos\theta t) \right]\\
n_\text{FID}(t)\simeq& e^{-(\Gamma_2+\Gamma_3)t}\cos B_0t\prod_{m=1}^M\cos\left({g_m\cos\theta t}\right)\notag\\ 
 & \left[1+\sum_{m=1}^M\epsilon_{1m}\tan\left(g_m\cos\theta t\right)\right]
\end{align}
where the relaxation and dephasing rates are given by
\begin{align}
\Gamma_1 =& \sum_{n=1}^N\frac{2\gamma_n g_{n}^2\sin^2\theta}{4\gamma_n^2+B_0^2},\label{eq:Gamma_1}\\
\Gamma_2 =& \frac{\Gamma_1}{2}+\Gamma_\phi \\
\Gamma_3 =& \sum_{m=1}^M \gamma_m \\
\Gamma_\phi =& \sum_{n=1}^N\frac{g_{n}^2\cos^2\theta}{2\gamma_n}.\label{eq:Gamma_phi}
\end{align}
Here $\epsilon _{1m}=\gamma _{m}/\left( g_{m}\cos \theta
\right) $ and $\epsilon_{2m}=g_m/B_0$
 are the small parameters of the perturbation theory.

The decoherence rates $\Gamma_1=1/T_1$, $\Gamma_2=1/T_2$ and $\Gamma_\phi$ are caused
by the fast fluctuators and the equations for them are consistent with Redfield 
results \cite{Slichter96}.
In the case of a single fast fluctuator, the decoherence rates for the echo
experiment can be directly connected to the noise power spectral density, i.e.,
$1/T_1=\sin^2\theta S_\text{RTN}(B_0)$, $1/T_\phi=\cos^2\theta S_\text{RTN}(0)$ 
and $1/T_2=1/2T_1 + 1/T_\phi$. 
$\Gamma_3$ is entirely due to the slow fluctuators.

\section{determination of model parameters from experimental data}
\label{sec:fit}

For purposes of data analysis, it is necessary for us to specify a not completely
general but yet still flexible model for the noise.  Let $d\left( \gamma
\right) =\sum_{k=1}^{K}\delta \left( \gamma -\gamma_{k}\right) $ be
the distribution of rates and take ${g}$ to be independent of ${k}$, i.e., $g_k=g$.
If there is a range of couplings then $g$ in the following formulas can be 
regarded as an appropriate average coupling.
    This equal-coupling-strength or single-coupling-strength assumption should not be
    a severe limitation of our model as long as the standard deviation in the distribution of $g_k$'s is small relative to $g$ itself, and to the width of 
    the distribution of $\gamma_k$'s.
For the specific case of fluctuating magnetic moments producing flux noise,  
the model is appropriate if the moments are all on the surface of the 
superconducting loop.  We will comment further on this below. 

We will assume a broad noise spectrum by taking 
\begin{align}
d\left(\gamma\right) =\begin{cases}
 \alpha \gamma ^{s-1}, &\text{for } \gamma_{\min }<\gamma <\gamma _{\max } \\
0, &\text{otherwise.}\end{cases}
\end{align} 
Here $\gamma_{\max}$ and $\gamma_{\min}$ are the upper and lower cuts of the 
fluctuators' switching rates.
For 1/f noise we must have $\gamma_{\min}>0$ and $\gamma_{\max}<\infty$ in order 
that the energy density of the noise be finite.
The power-law assumption is often useful for analyzing experimental data, though the 
method used to solve the model itself is capable of treating arbitrary distributions.
Note $s=0$ gives 1/f noise.  

Under those assumptions, the pulsed signals are given by
\begin{align}
     n_\text{ER}(t) \simeq& e^{-\Gamma_{1}t}\label{eq:ER}\\
n_\text{SE}(t) \simeq&  e^{-(\Gamma_2+\Gamma_3)t}\left[1 + \frac{\Gamma_3}{\gamma_c}\sin(\gamma_c t) \right] \label{eq:Echo_IC}\\
n_\text{FID}(t) \simeq& e^{-(\Gamma_2+\Gamma_3) t}\cos B_0t\cos^{M}(\gamma_c t) 
    \left[1+\frac{\Gamma_3}{\gamma_c}\tan(\gamma_c t)\right]\label{eq:FID_IC} 
\end{align}
where $\gamma_c=g\cos\theta$ is the critical coupling strength. 
The task of data analysis is then to determine the five intrinsic parameters $g$, 
$\alpha$, $s$, $\gamma_{\min}$, and $\gamma _{\max}$ from observations of the 
pulsed signals $n_\text{ER}$, $n_\text{FID}$ and $n_\text{SE}$.
The formulas show that the ER and SE signals at different working points alone are 
enough to fully determine all the five parameters, at least in principle. 
The FID data provide consistency checks and, crucially, to find the number of  
slow fluctuators at various working points.
We have analyzed data from Ref. \cite{Yoshihara06,Kakuyanagi07} and all results are 
listed in Table \ref{tab:results}.

For the ease of analysis, it is convenient to define 
\begin{align}
\Phi _\text{SE}\left( t,\theta \right) &=\frac{n_\text{SE}\left( \theta \right) }{
n_\text{SE}\left( \theta =\pi /2\right) } \notag\\
&=e^{-\Gamma _{3}t}\left[ 1+\frac{\Gamma _{3}}{\gamma _{c}}\sin \gamma _{c}t
\right],  \label{eq:phiSE}
\end{align}
where $\Phi _\text{SE}\left( t,\theta =0\right)$ correspond to the `phase-memory functional' defined by other authors \cite{Galperin06,Galperin07}.
Note $\Gamma_1$ has only weak dependence on the working point $\theta$, 
thus $\Gamma_2(\theta)\simeq \Gamma_1(\theta=\pi/2)/2$, 
and it drops out in Eq. \ref{eq:phiSE}. 

Similarly, we define $\Phi_\text{FID}$ or the FID signal
\begin{align}
\Phi_\text{FID}\left( t,\theta \right) &=\frac{n_\text{FID}\left( \theta \right) }{
n_\text{FID}\left( \theta =\pi /2\right) } \notag\\
&=e^{-\Gamma _{3}t}\left[ 1+\frac{\Gamma _{3}}{\gamma _{c}}\tan \gamma _{c}t \right]
\cos^M(\gamma_c t).  \label{eq:phiFID}
\end{align}
It is important to note that $\Phi_\text{FID}$ has explicit dependence on the number of slow RTN fluctuators $M$.

We note the scaling parameter $s$ has significant effect on the working point 
dependence of the decoherence rates, especially for $\Gamma_3$.
In the case of 1/f noise, $s=0$ and we have
\begin{align}
\Gamma_1 \simeq& \frac{\alpha g^2\sin^2\theta}{B_0}\tan^{-1}\left(
\frac{2\gamma_{\max}}{B_0}\right)\label{eq:g1}\\
\Gamma_2 \simeq& \frac{\Gamma_1}{2}+\frac{\alpha}{2}\gamma_c \label{eq:g2}\\
\Gamma_3 \simeq& \alpha(\gamma_c-\gamma_{\min}). \label{eq:g3}
\end{align}
Note $\Gamma_3$ has linear relationship to $\gamma_c$ in this case. 

If $s= 1$, 
\begin{align}
\Gamma_1 \simeq& \frac{\alpha g^2\sin^2\theta}{4}\log\left(\frac{B_0^2+4\gamma_{\max}^2}{B_0^2+4\gamma_c^2}\right)\\
\Gamma_2 \simeq& \frac{\Gamma_1}{2}+\frac{\alpha\gamma_c^2}{2}\log\frac{\gamma_{\max}}{\gamma_c}\\
\Gamma_3 \simeq& \frac{\alpha}{2}(\gamma_c^{2}-\gamma_{\min}^{2}).
\end{align}

In general,
\begin{align}
\Gamma_1 \simeq& \frac{2\alpha\sin^2\theta g^2 \gamma^{s+1}}{B_0^2(s+1)} {_2F_1}
\left(1,\frac{s+1}{2};\frac{s+1}{2}+1;-\frac{4\gamma^2}{B_0^2} \right)\\
\Gamma_2 \simeq& \frac{\Gamma_1}{2}+\frac{\alpha\gamma_c^2}{2(s-1)}\left({\gamma_{\max}}^{s-1}-{\gamma_c}^{s-1}\right)\\
\Gamma_3 \simeq& \frac{\alpha}{s+1}(\gamma_c^{s+1}-\gamma_{\min}^{s+1}).\label{eq:G3_general}
\end{align}
Here $_2F_1$ is the hypergeometric function. 

In flux qubit experiments, the working point is experimentally tunable by varying 
the applied flux and $\gamma_c = g\cos\theta$ changes accordingly. 
Thus a plot of $\Gamma_3$ versus $\gamma_c$ would unambiguously determine the 
distribution of the fluctuators $d(\gamma)$.

\subsection{coupling constant $g$}

\begin {figure}[tbp]
    \centering\includegraphics*[width=\linewidth]{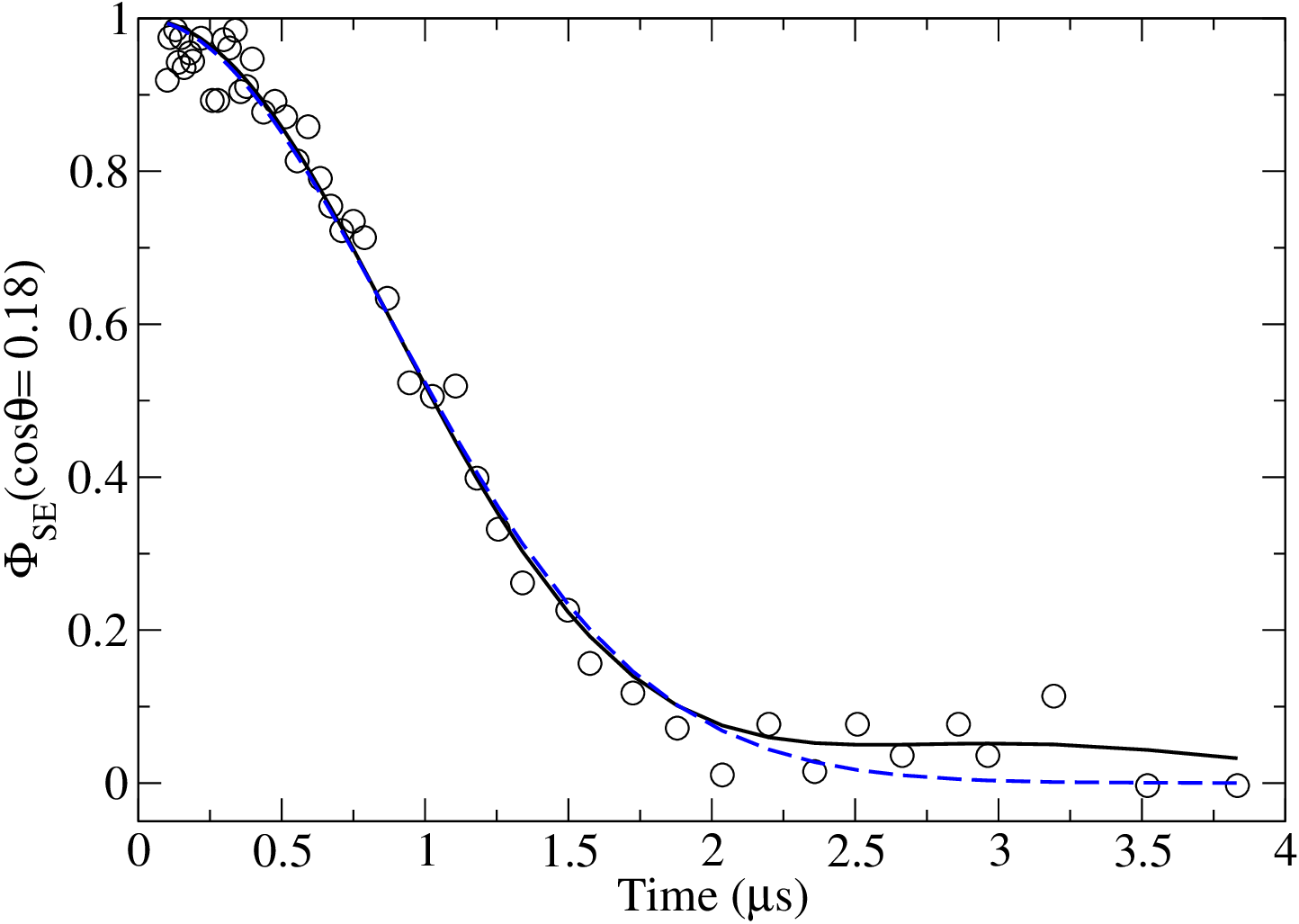}
    \caption{(Color Online) Echo phase memory functional  $\Phi_{SE}$ data in Fig.4A 
of Ref.\cite{Yoshihara06}. We fit the $36$ data points (open circle) 
to Eq.\ref{eq:phiSE} (black solid line) and 
Gaussian model $\Phi_{SE}^G=e^{-\Gamma_G^2t^2}$ (blue dashed line) respectively. }
\label{fig:echo_4A}
\end{figure}

To extract the coupling constant $g$,  we fit the phase memory functional to Eq. 
\ref{eq:phiSE}, as seen in Fig. \ref{fig:echo_4A}. 
Thus for each working point $\theta$, we can extract two numbers from the 
fitting, i.e. $\Gamma_3(\theta)$ and $\gamma_c(\theta)$. 
For example, the data in Fig. \ref{fig:echo_4A} were taken at 
working point $\cos\theta=0.18$, and the corresponding $\Gamma_3 =0.99$ MHz and 
$\gamma_c=2.1$ MHz.

    In Ref. \cite{Yoshihara06,Kakuyanagi07}, the same data is fitted to a 
    Gaussian noise model where the Gaussian flux fluctuation assumes a 1/f 
    noise spectral density, i.e., $S_\phi(\omega) = A/\omega$. 
    In this case, the phase memory functional takes the Gaussian form 
    $\Phi_\text{SE}^G=e^{-\Gamma_G^2t^2}$.
    As seen in Fig. \ref{fig:echo_4A}, both models fit the data well and it is 
    is unclear which model is better.
    Non-Gaussian behavior manifests itself unambiguously with `plateaus' in the
    phase memory functional $\Phi_\text{SE}$ \cite{Galperin06,Galperin07}.
    The rise in the longer time in Fig. \ref{fig:echo_4A} could be the onset 
    of such `plateaus'. 
    A cleaner sample with fewer surface spins (smaller $\alpha$) would help to 
    make the `plateaus' more visible, which would then distinguish the 
    present model from the Gaussian model \cite{Zhou10PRA}.

\begin {figure}[tbp]
    \includegraphics*[width=\linewidth]{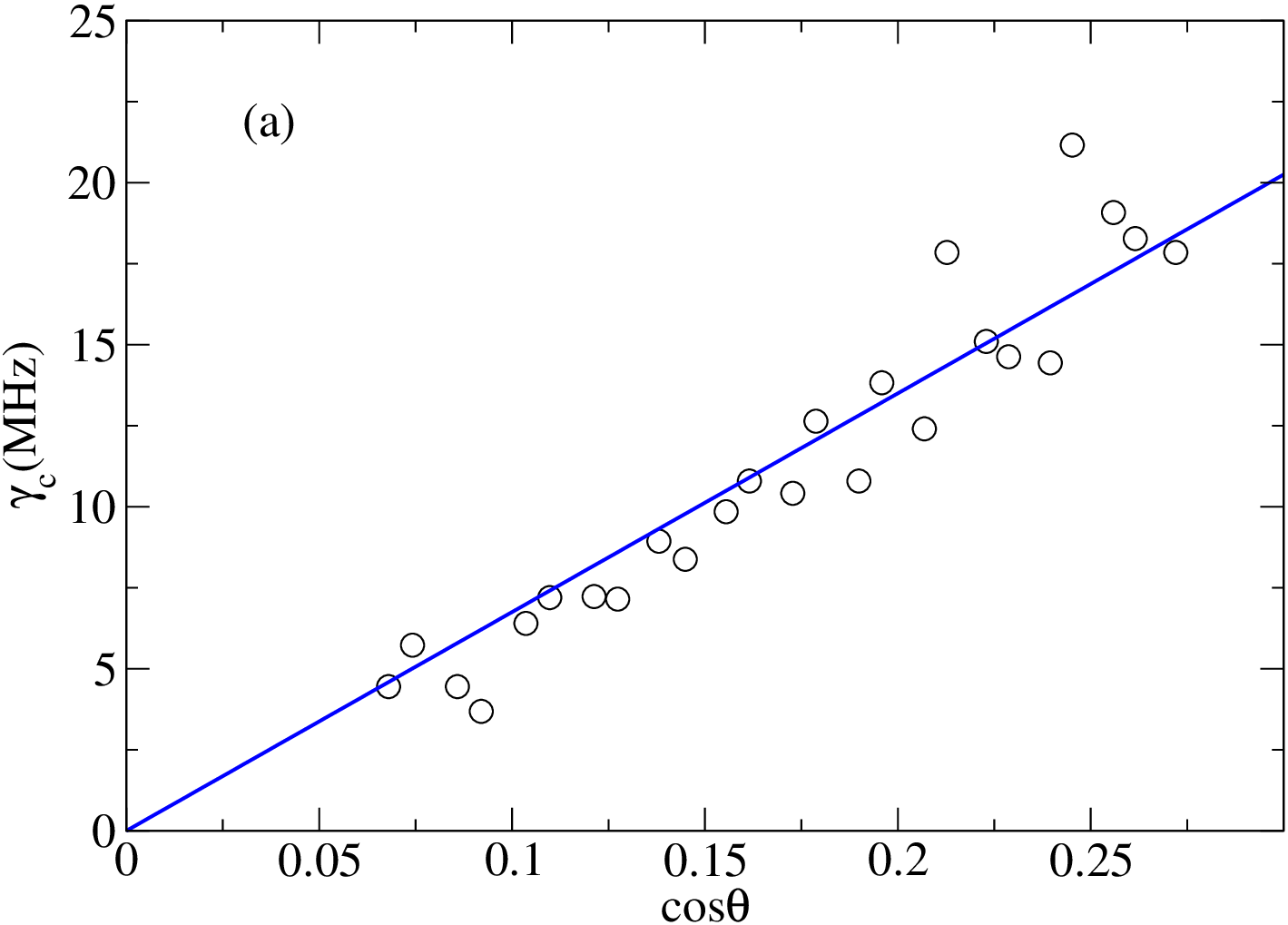}
    \includegraphics*[width=\linewidth]{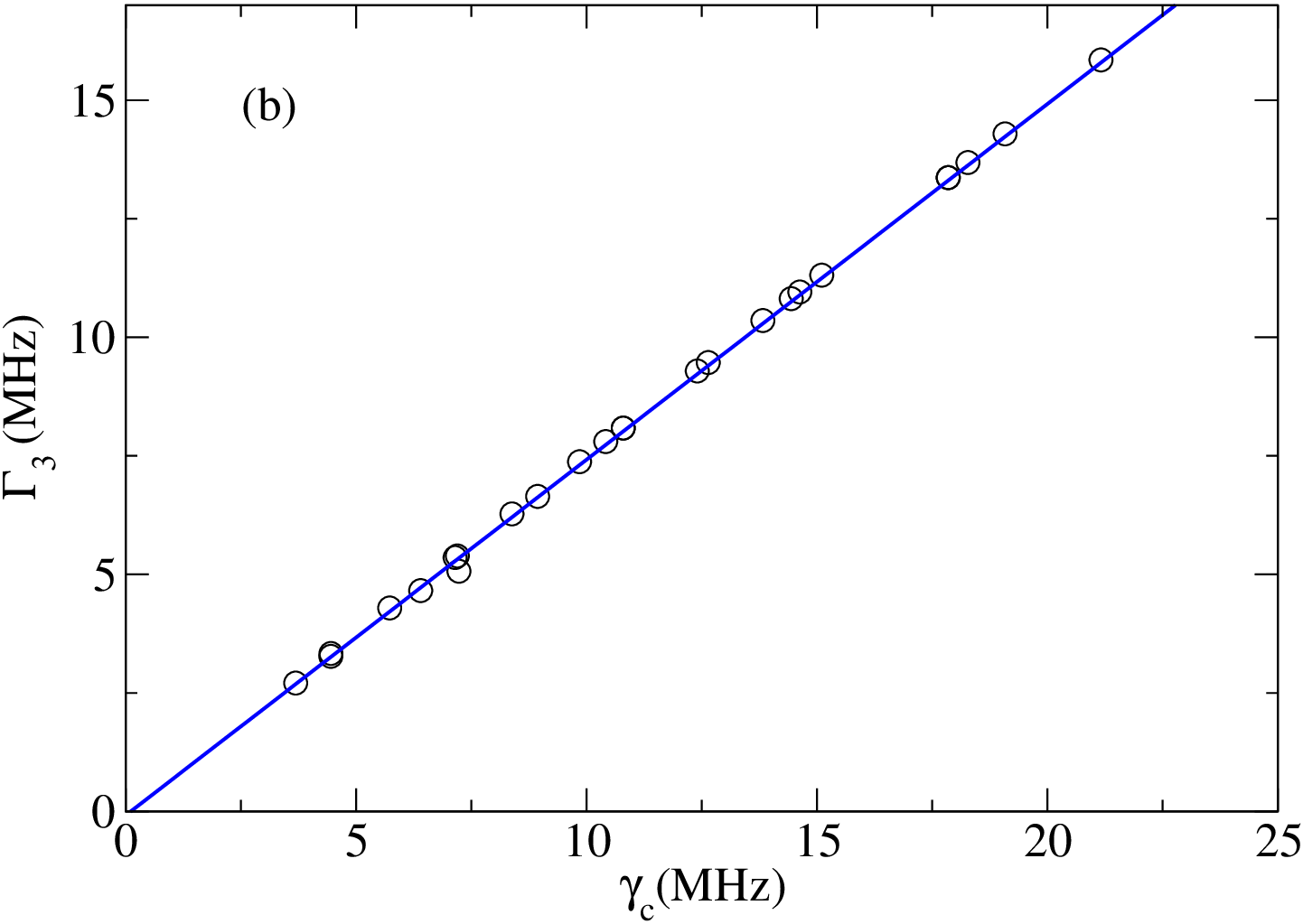}
    \caption{Fitting of data from Ref.\cite{Kakuyanagi07}. Both $\gamma_c$ and $\Gamma_3$ are fitted from $\Phi_{SE}$ at various working point. 
(a) Critical rate $\gamma_c$ versus the working point $\cos\theta$. 
The slope is the coupling constant $g$.
(b)  Linear regression to $\Gamma_3=\alpha(\gamma_c-\gamma_{\min})$.
}\label{fig:kaku_fit}
\end{figure}

With data at different working point $\theta$, we can plot $\gamma_c$
versus $\cos\theta$.  The coupling constant $g$ is the slope, as seen 
in Fig.\ref{fig:kaku_fit}(a) \footnote{In the data analysis, we first extract
$\Gamma_G$ from Ref. \cite{Yoshihara06,Kakuyanagi07}, then reproduce
$\Phi^G_\text{SE}$ from the Gaussian formula. Given $\Phi^G_\text{SE}$ fits 
the real experimental data well, we fit 
the reproduced $\Phi^G_\text{SE}$ to Eq. \ref{eq:phiSE} to obtain $\Gamma_3$
and $\gamma_c$ at different working points. }.
    Similar data analysis for Ref. \cite{Yoshihara06} has been carried out 
    in Ref. \cite{Zhou10PRA}.

\subsection{noise intensity $\alpha$, noise index $s$ and lower cut $\gamma_{\min}$}

The functional form of $\Gamma_3$ allows us to determine $\alpha$, $s$ and $\gamma_{\min}$, as seen in Eq. \ref{eq:G3_general}.

Fitting the data from Ref. \cite{Kakuyanagi07} we get clean linear dependence 
of $\gamma_c$, as seen in Fig. \ref{fig:kaku_fit}(b). 
A similar result has been obtained in Ref. \cite{Zhou10PRA} for the data 
from Ref. \cite{Yoshihara06}.
This is a sign of 1/f noise in the environment ($s=0$). 
$\gamma_{\min}$ is the intercept of the linear fit. 
Unfortunately, its accuracy depends strongly on the quality of the data at 
low $\gamma_c$, and the data are lacking in that region. 
Hence there is considerable uncertainty in the fitted value of $\gamma_{\min}$.
The noise intensity $\alpha$ can be retrieved from the slope, in the case of 
1/f noise, as seen in Eq. \ref{eq:g3}.

    If $s$ is taken as a fitting parameter as well, we get $s = -0.005$ for
    Ref. \cite{Kakuyanagi07} and $s=-0.07$ for Ref. \cite{Yoshihara06}.
    Thus the data is consistent with 1/f noise and we adopted the linear fit
    as in Eq. \ref{eq:g3}.
    When the noise power spectral density deviates farther from $1/f$, 
    there are indications that as $s$ decreases from $0$, 
    the pure dephasing time $T_\phi$ also decreases \cite{Anton2011}.
 
\subsection{upper cutoff $\gamma_{\max}$}

For general $s$, $\Gamma_1$ can be expressed in terms of hypergeometric function in 
$s$. In the case of $1/f$ noise, $s=0$ and we have Eq. \ref{eq:g1}.
Since $\gamma_{\max}$ is the only unknown in the equation ($\Gamma_1$ is 
experimentally measurable and the other quantities can be derived from the experiment),
 $\gamma_{\max}$ can be extracted, at least in principle.
But for the data in Ref. \cite{Yoshihara06, Kakuyanagi07}, 
we are unable to back out $\gamma_{\max}$. 
One finds $\tan^{-1}(2\gamma_{\max}/B_0)$ has to be greater than $\pi/2$ to validate the equation.

The lack of experimental accuracy might not help for this self-inconsistency. 
One possible remedy is that there is some other source of high frequency noise, other than RTN, to cause relaxation.
    Thus the experimentally observed $\Gamma_1$, as denoted by 
    $\Gamma_1^{(ex)}$ in Table \ref{tab:results} is actually greater than 
    $\Gamma_1^{(th)}=\pi\alpha g^2/2B_0$ which only includes the energy
    relaxation due to the fluctuators.
    Here $\Gamma_1^{(th)}$ is defined with $\gamma_\text{max}=\infty$ and 
    $\theta = \pi/2$ for Eq. \ref{eq:g1}.

\begin{table}[tb]
\begin{tabular*}{\linewidth}{@{\extracolsep{\fill}}|c|c|c|}
\hline 
 & A & B \tabularnewline
\hline 
$\Delta/h$ (GHz) & $5.445$ & $3.9$\tabularnewline
\hline 
$\varepsilon/h$ (GHz) & $0\sim1$ & $0\sim1.1$\tabularnewline
\hline 
$I_p$ (nA) & $160$ & $370$ \tabularnewline
\hline
$r$ ($\mu$m)& $1$& $2$\tabularnewline
\hline 
$\Gamma_1^{(ex)}$ (MHz) & $0.65$ & $7.1$\tabularnewline
\hline 
\hline 
$g/h$ (MHz) & $9.5(0.2)$  & $68(2)$ \tabularnewline
\hline 
$\gamma_{\min}$ (MHz) & $0.05(0.01)$ & $0.11 (0.05)$  \tabularnewline
\hline 
$\alpha$ & $0.77(0.006)$& $0.754(0.003)$\tabularnewline
\hline 
$\Gamma_1^{(th)}$ (MHz) & $0.02(0.01)$& $1.4(0.1)$\tabularnewline
\hline 
$m$ ($\mu_B$)& $1.7(0.03)\times 10^3$& $5.2(0.2)\times 10^3$\tabularnewline
\hline 
$S_\Phi(\omega=1Hz)$ ($\Phi_0^2/Hz$) & $3.5(0.2)\times 10^{-11}$& $3.3(0.2)\times 10^{-10}$\tabularnewline
\hline 
\end{tabular*}
\caption{Noise characteristics extracted from Ref.\cite{Yoshihara06} (column A) and Ref.\cite{Kakuyanagi07} (column B).
The numbers in the top portion are experimental data while the ones in the lower portion are derived parameters from the model.
The numbers in the parenthesis are standard deviations from linear regression. 
}\label{tab:results}
\end{table}

\subsection{number of fluctuators $K$}

Since most of the parameters in the model can be derived from the experimental data 
 for $n_\text{SE}$ and $n_\text{ER}$, 
we may use the data for $n_\text{FID}$ to get a value for
$M(\theta)$, the number of slow fluctuators. 
 As seen in Eq. \ref{eq:phiFID}, the FID signal has explicit dependence on $M$.
It is easiest to analyze this using the
logarithm of the phase memory functional $\mathcal{K}_E(t)$. 

For the Echo signal, the logarithm of the phase memory functional $\mathcal{K}_E(t)$ can be expanded in terms of $\gamma_ct\ll 1$ and we have
\begin{equation}\label{eq:K_E}
\begin{aligned}
\mathcal{K}_E(t)\equiv& -\log \Phi_{SE}\\
        \simeq&\left\{\begin{aligned}
&\Gamma_3\gamma_c^2t^3/6, \quad & \gamma_c t \gg \Gamma_3/\gamma_c,\\
&\Gamma_3^2t^2/2, \quad & \gamma_c t\ll \Gamma_3/\gamma_c.
\end{aligned} \right.
\end{aligned}
\end{equation}

Similarly, we define $\mathcal{K}_F$ for the envelope of FID signal. In the limit of $\gamma_c t\ll 1$,
\begin{equation}\label{eq:K_F}
\begin{aligned}
\mathcal{K}_F(t) \equiv& -\log\Phi_\text{FID},\\
        & \simeq \frac{M\gamma_c^2+\Gamma_3^2}{2}t^2.
\end{aligned}
\end{equation}

Note at small times ($\gamma_ct\ll\Gamma_3/\gamma_c$), both $\mathcal{K}_E(t)$ and 
$\mathcal{K}_F(t)$ are quadratic in time. If the waiting time for the first plateau 
is too long comparing to the damping time $\tau_{D}\simeq1/\Gamma_3$, 
both phase memory functionals will assume Gaussian shape. In the experiments 
\cite{Yoshihara06,Kakuyanagi07}, the authors fit to the quadratic terms in Eqs. 
\ref{eq:K_E} and \ref{eq:K_F}. What they called $\Gamma_{\phi E}^g$ and $\Gamma_{\phi F}^g$ correspond to $\Gamma_3/\sqrt{2}$, and $\sqrt{(M\gamma_c^2+\Gamma_3^2)/2}$. 

In Ref. \cite{Yoshihara06}, $\Gamma_{\phi F}^g\simeq 8\Gamma_{\phi E}^g$. 
This linear dependence is expected as long as the qubit is not operated extremely close to the 
optimal point $\theta=\pi/2$, as can be seen from Eq. \ref{eq:g3}.
Thus the number of slow fluctuators $M\simeq64\Gamma_3^2/\gamma_c^2$ is of the order 
$10$ for the working points in the experiment. To be more specific,
\begin{align}
M\sim 64\times\alpha^2 \sim 38.
\end{align}
It should be evident that this is a rough estimate.  
However, it is unlikely to be off by order of magnitude and 
we assert that $M < 100$.

\subsection{effective magnetic moment and power spectrum}

As a consistency check, we find the magnetic moment associated with the
fluctuators and the total spectral density. The change in flux due to spin in the 
SQUID loop is 
\begin{equation}
\Delta \Phi = \frac{\mu_0}{r} m
\end{equation}
where $m$ is the effective magnetic moment of the spin, $\mu_0$ is the magnetic 
constant and $r$ is the radius of the loop.

For flux qubit, we have
\begin{equation}
\Delta \Phi = \frac{g}{2I_p}, \label{eq:mag}
\end{equation}
where $I_p$ is persistent current along the qubit loop. Thus
\begin{align}
m = \frac{r g}{2\mu_0 I_p}.
\end{align}

The noise power spectrum density is 
\begin{align}
S_{1/f}(\omega) =& \int_{\gamma_{\min}}^{\gamma_{\max}}S_{\text{RTN}}(\omega)d(\gamma)d\gamma\notag \\
\simeq& \frac{g^2}{\pi \omega}\left[\tan^{-1}\left(\frac{2\gamma_{\max}}{\omega}\right)-\tan^{-1}\left( \frac{2\gamma_{\min}}{\omega}\right) \right]\notag \\
\simeq & \frac{\alpha g^2}{2\omega}
\end{align}
With Eq. \ref{eq:mag}, the noise power spectrum density in terms of flux is 
\begin{align}
S_{\Phi}(\omega) = S_{1/f}(\omega)/4I_p^2.
\end{align}

These derived results are listed in Table \ref{tab:results}. 

\section{conclusion}
\label{sec:conc}
We have given a method for extracting the properties of the two-state 
fluctuators that cause decoherence in superconducting qubits.  
This method applies when the working point of the qubit can be varied, 
as is possible in flux qubit set-ups.  
The shape and strength of the noise spectrum can be determined
from qubit measurements, and an estimate of the total number of active slow
fluctuators can be obtained.  
We analyze two experiments and find that the number
of slow fluctuators is surprisingly small, less than $100$. 
If we assume that the noise arises from magnetic clusters 
on the surface of the superconducting loop, then the size of the magnetic 
moment of the clusters is quite large, of the order of $1000$ to $5000\mu_B$.  

These results appear to be rather surprising.  
However, a recent analysis \cite{Kechedzhi11} of noise measurements on 
dc SQUID inductance \cite{Sendelbach09} suggests that the 
predominant noise sources are large magnetic clusters and that 
such clusters would give rise to 1/f-like noise.  
This gives rise to a consistent picture of two quite 
different qubits analyzed in two quite different ways.

    The quasi-Hamiltonian method is applicable to more complex systems as well,
    such as interacting two-qubit systems \cite{De11}.
An extension of the current work would be to examine the  more recent 
experiment where the dephasing of two inductively coupled flux qubits 
are studied \cite{Yoshihara2010}.

\begin{acknowledgments}
We thank Robert McDermott, J. S. Tsai and K. Kechedzhi for helpful discussions 
and Zhigeng Geng for teaching the authors to use R 
(the programming language and software environment for statistical computing 
and graphics).
This work is supported by the DARPA/MTO QuEST program through a grant from AFOSR.
\end{acknowledgments}

\bibliography{/home/nos/projects/library}       

\end{document}